\documentclass[aps,prl,twocolumn,letterpaper,superscriptaddress,10pt]{revtex4-2}
\usepackage{amssymb,amsthm,amsmath,amsfonts}
\usepackage{graphicx,ulem,soul,enumerate}
\usepackage[pdftex,dvipsnames,usenames]{xcolor}
\usepackage[colorlinks=true,urlcolor=blue,citecolor=blue,linkcolor=blue]{hyperref}

\begin{document}

\title{Subdiffusion via Disordered Quantum Walks}

\author{Andrea Geraldi}
\email{andrea.geraldi@uniroma1.it}
\affiliation{Dipartimento di Fisica, Sapienza Universit\`{a} di Roma, Piazzale Aldo Moro, 5, I-00185 Roma, Italy}

\author{Syamsundar De}
\email{syamsundar.de@upb.de}
\affiliation{Integrated Quantum Optics Group, Applied Physics, Paderborn University, 33098 Paderborn, Germany}

\author{Alessandro Laneve}
\affiliation{Dipartimento di Fisica, Sapienza Universit\`{a} di Roma, Piazzale Aldo Moro, 5, I-00185 Roma, Italy}

\author{Sonja Barkhofen}
\affiliation{Integrated Quantum Optics Group, Applied Physics, Paderborn University, 33098 Paderborn, Germany}

\author{Jan Sperling}
\affiliation{Integrated Quantum Optics Group, Applied Physics, Paderborn University, 33098 Paderborn, Germany}

\author{Paolo Mataloni}
\affiliation{Dipartimento di Fisica, Sapienza Universit\`{a} di Roma, Piazzale Aldo Moro, 5, I-00185 Roma, Italy}

\author{Christine Silberhorn}
\affiliation{Integrated Quantum Optics Group, Applied Physics, Paderborn University, 33098 Paderborn, Germany}

\date{\today}


\begin{abstract}
	Transport phenomena play a crucial role in modern physics and applied sciences.
	Examples include the dissipation of energy across a large system, the distribution of quantum information in optical networks, and the timely modeling of spreading diseases.
	In this work, we experimentally prove the feasibility of disordered quantum walks to realize a quantum simulator that is able to model general subdiffusive phenomena, exhibiting a sublinear spreading in space over time.
	Our experiment simulates such phenomena by means of a finely controlled insertion of various levels of disorder during the evolution of the walker, enabled by the unique flexibility of our setup.
	This allows us to explore the full range of subdiffusive behaviors, ranging from anomalous Anderson localization to normal diffusion.
\end{abstract}

\maketitle

\paragraph*{Introduction.---}

	Transport phenomena are ubiquitous in physics, often in connection with the prominent heat equation.
	Such phenomena are prime examples for normal diffusion processes in which the variance, quantifying the spatial spread of the system's distribution, increases linearly with time.
	Nevertheless, it is actually quite common to find natural processes featured by a dynamics which does not follow such a simple relation.
	Rather, these systems are characterized by a distribution that broadens according to a nonlinear power law \cite{metzler2004restaurant, metzler2014anomalous}, a behavior referred to as anomalous diffusion.
	In particular, a sublinear relation between variance and time, i.e., subdiffusion, can be frequently observed in nature, such as in biological processes \cite{bronstein2009transient, coker2019controlling, regner2013anomalous, golan2017resolving}, wave propagation and scattering \cite{marano2005ray, toivonen2018anomalous}, the movement of charge carriers in amorphous semiconductors \cite{scher1975anomalous}, disordered media \cite{havlin2002diffusion}, and many-body localization transitions \cite{agarwal2015anomalous}.
	Subdiffusion even applies to certain economic models \cite{masoliver2003continuous}.

	Because of this vast range of applications and its fundamental importance, a manifold of attempts have been made during recent years to uncover the underlying physical mechanisms that lead to anomalous diffusion.
	Such theoretical models rely on a variety of physically motivated and more abstract approaches, such as fractal theory \cite{chen2010anomalous, ben2000diffusion}, fractional Brownian motion \cite{jeon2010fractional}, and continuous-time random walks \cite{montroll1965random, abad2010reaction}.
	Consequently, the possibility to simulate all kinds of anomalous diffusive behaviors in one platform---and in a tunable manner---can not only lead to significant insights into complex mathematical models but also enables us to study a plethora of processes in nature.
	Here, we show that such a simulation task can, indeed, be realized by means of disordered quantum walks (QWs).

	QWs---the counterpart to classical random walks that exploit coherent superpositions---serve as a promising framework to implement simulation protocols since they provide a general model for the propagation of quantum particles \cite{aharonov1993quantum, venegas2012quantum, kempe2003quantum}.
	For example, QWs have been used to study transport phenomena in biomolecules \cite{Plenio2008biomolecules}, evolution in solid-state systems \cite{christandl2004spin}, formation of molecular states \cite{albrecht2912molecular}, topological invariants \cite{ramasesh2017direct, barkhofen2017measuring}, and edge states \cite{xiao2017observation, nitsche2019eigenvalue}.
	However, basic QWs are characterized by a spread which grows quadratically in time.
	This superdiffusive broadening is referred to as a ballistic diffusion.
	Moreover, by actively influencing the evolution of the walker, the functional dependency of the broadening can be altered, e.g., for reproducing the classical normal diffusion of incoherent processes.
	For instance, static disorder leads to Anderson localization, arising from the interaction between the coherent quantum walker and a disordered environment \cite{crespi2013anderson, schreiber2011decoherence}.

	Recently, the continuous transition from ballistic behavior to normal diffusion has been experimentally demonstrated in a QW through the implementation of time and space inhomogeneous evolution pattern, according to a so-called $p$-diluted model \cite{geraldi2019superdiffusion}.
    This approach fundamentally proves that superdiffusion is achievable by introducing inhomogeneities in the QW's evolution.
    Nevertheless, since that experiment was based on a spatial scheme that is comparably hard to scale, it was only possible to investigate the superlinear regime.
	Therefore, the less accessible entirety of the subdiffusive domain remains largely unexplored.
	Among other reasons, a lack of a fitting experimental platform hindered such a realization to date since it has to be scalable, dynamically reconfigurable, and compatible to the introduction of controlled disorder in spatial and temporal domain in order to realize advanced disorder models.

	In this paper, we experimentally demonstrate that the conceptual idea of $p$-diluted disorder can be critically extended to encompass the subdiffusive regime as well.
	In contrast to earlier implementations of the superdiffusive domain, our experiment uses a highly flexible time-multiplexing scheme to resolve the open problem of simulating subdiffusion processes.
	By controlling disorder in the spatial degree of freedom (here, time bins) and in time (here, number of steps), we show that it is possible to realize any sublinear propagation regime---ranging from statically disordered QWs, giving birth to Anderson localization, to completely disordered QWs, corresponding to normal diffusion---that is achieved by implementing different disorder levels.

\paragraph*{Theoretical model.---}

	A broadly applicable model for diffusion and general transport processes can be formulated in terms of the partial differential equation
	\begin{equation}
		\label{eq:diffeq}
		0=\partial_t P(x,t)+\mathcal L P(x,t),
	\end{equation}
	in which $P(x,t)$ represents a space-time dependent probability distribution and $\mathcal L$ is a potentially time-dependent differential operator in the spatial degree of freedom $x$.
	For example, $\mathcal L\propto-\partial_x^2$ describes the heat equation that results in normal diffusion.
	For a large family of randomized media, the asymptotic solution for large times $t$ reads
	\begin{equation}
		\label{eq:spatial}
		P(x,t)\propto\exp\left(-\left|\frac{ax}{\sigma(t)}\right|^b\right),
	\end{equation}
	where $b$ describes the type of the exponential decay and $a$ is a scaling factor.
	Furthermore, $\sigma(t)$ is the standard deviation with the characteristic power law behavior,
	\begin{equation}
		\label{eq:temporal}
		\sigma(t)=ct^{d},
	\end{equation}
	where $2d$ determines the spread of the variance over time and $c$ is another scaling factor.
	See, e.g., Ref. \cite{giona1992fractional} for a thorough derivation of this model.
	For example, for $b=2$ and $2d=1$, we get a Gaussian distribution in space with a linear increase of the variance.
	And the parameters $b=1$ and $2d=0$ result in Anderson localization as a result of the static disorder.
	Here, we aim at exploring the theoretically predicted intermediate regime, $1< b< 2$, with a subdiffusive behavior, $0<d<1/2$.

	As established before, discrete QWs have shown their ability to simulate certain diffusion regimes, such as superdiffusive power laws \cite{geraldi2019superdiffusion}, in the continuous limit of many steps and positions.
	The walker on a line is described by the coherent superposition state, $|\psi(t)\rangle=\sum_{x}(\psi_0(x,t)|x\rangle\otimes|0\rangle+\psi_1(x,t)|x\rangle\otimes|1\rangle)$, where $\{|0\rangle,|1\rangle\}$ represents the quantum coin of the walker.
	For the resulting probability distribution, we trace over this internal degree of freedom, $P(x,t)=|\psi_0(x,t)|^2+|\psi_1(x,t)|^2$.
	The QW evolves by means of the action of two operators, the coin operator $\hat{C}(t)$ and the step operator $\hat{S}$.
	The coherent coin toss is given by the unitary map $\hat{C}(t) = \sum_{x}|x\rangle\langle x|\otimes\hat{C}(x,t)$, which can vary with positions and times. 
	The step operator, $\hat{S} = \sum_{x}\left(|x-1\rangle\langle x|\otimes |0\rangle\langle 0|+|x+1\rangle\langle x|\otimes|1\rangle\langle 1|\right)$, then coherently propagates the walker in the two directions, depending on the coin.
	Thus, the evolution of the full quantum system can be expressed by $|\psi(t+1)\rangle = \hat{S} \hat{C}(t) |\psi(t)\rangle$.
	
	It turns out to be convenient to model different anomalous diffusion regimes with a corresponding degrees of static and dynamic disorder in the choice of the space-time dependent coin.
	The degree of the dynamic variation is determined by a parameter $p$, resulting in the notion of $p$-diluted disorder  \cite{geraldi2019superdiffusion}.
	In general, the coin operator is not homogeneous with respect to position and step and different constraints can be imposed \cite{yin2008quantum}.
	For instance, the coin operator can be inhomogeneuous in space, but static in time, $\hat{C}(x,t) = \hat{C}(x)$, leading to Anderson localization ($b=1$) around the starting position $x=0$ \cite{anderson1958absence}, which is a static effect ($2d=0$).
	Now, this static disorder can be perturbed in the $p$-diluted model to approximate different differential operators $\mathcal L$ in Eq. \eqref{eq:diffeq} for different physical scenarios.
	This perturbation consists of the independent and random choice of time-dependent coin configurations according to the percentage of dynamic disorder $p$,
	\begin{equation}
	    \label{eq:randomcoin}
	\begin{split}
	    \hat{C}(x,t)=\left\lbrace\begin{array}{ll}
		    \hat{C}(x,t) & \text{with probability }p,
		\\
		    \hat{C}(x) & \text{with probability }1-p,
		\end{array}\right.
	\end{split}
	\end{equation}
	which introduces an inhomogenitiy in time.
	Specifically, $p=0$ yields Anderson localization ($b=1$ and $2d=0$), and $p=1$ results in a completely disordered QW ($b=2$ and $2d=1$).
	Most importantly, the region $0<p<1$ should theoretically enable us to control our QW in such a way that it explores the full intermediate range of exponential spatial decays, $1<b<2$ in Eq. \eqref{eq:spatial}, with sublinear temporal spreads, $0<2d<1$ in Eq. \eqref{eq:temporal}.
	See the Supplemental Material (SM) for technical details and the connection to transport processes.

\paragraph*{Experimental implementation.---}

	To demonstrate such subdiffusive phenomena, our QW experiment relies on the well-established time-multiplexing scheme based on an unbalanced Mach-Zehnder interferometer with a feedback loop \cite{schreiber2010photons, schreiber2011decoherence, Nitsche2018recurrence}.
	Our time-multiplexing scheme provides high resource efficiency, long-lasting stability, and homogeneity, which we exploit for the realization of QWs over a sufficiently large number of steps that is necessary for clearly distinguishing the signatures of the subdiffusive behavior.
	This was not possible through the previous spatial implementation of $p$-diluted model for super-diffusion \cite{geraldi2019superdiffusion}.

	In the present scheme, the position degree of freedom is encoded in the arrival time bin of a weak coherent laser pulse at the single-photon level that acts as the walker while the coin information is embedded in the polarization, $|H\rangle= |0\rangle$ and $|V\rangle = |1\rangle$.
	An unbalanced interferometer, acting as a delay line by introducing a well-defined delay between the two polarizations, realizes the step operation.
	A significant step forward in comparison to the previous time-multiplexing setups is the introduction of a fast-switching electro-optic modulator (EOM) in the feedback loop that enables the dynamical control over coin operation via polarization rotations, without introducing high additional losses.
	This position and step dependent coin operation is harnessed for the implementation of $p$-diluted disorder that is central to the realization of subdiffusive dynamics. 
	Additionally, we employ two EOMs in the interferometer arms allowing for deterministic routing of the walker either into the feedback loop or towards the detection unit that measures photon's arrival time (position) and polarization (coin) degrees of freedom.
	For further experimental details, see the SM.

\paragraph*{Results.---}

    Our goal is to explore subdiffusive dynamics by carefully studying the behavior of the walker as a function of the disorder level. 
    In theory, for a sufficiently large number of steps, our measured data for a discrete QW should approach the continuous model of subdiffusion.
	For a given amount of disorder $p$, several coin configurations can be obtained because of the randomness in the choice of the coin [Eq. \eqref{eq:randomcoin}].
	We refer to each configuration as a coin map.
	Relevant quantities can be extracted from the walker's output probability distribution $P(x,t)$ after averaging it over many realizations, performed with the given disorder value $p$.
	We experimentally implemented $400$ coin maps for each disorder scenario under study,
	\begin{equation}
	    \label{eq:ps}
	    p\in\{0.0,0.1,0.2,0.3,0.5,1.0\},
	\end{equation}
	and the resulting average probability distribution has been measured for step numbers
	\begin{equation}
	    \label{eq:ts}
	    t\in\{5,8,11,14,17,20\}.
	\end{equation}
	For each of the $2\,400$ coin maps, we create a statically disordered coin configuration which are then perturbed according to the given level $p$ of disorder.
	By randomly choosing the starting static disorder, it is assured that the final results do not depend on a particular static configuration but only on the disorder level $p$.

	Our measured data enables us to analyze both the spatial characterization in terms of $P$ with a given exponential behavior, $1<b<2$, and the temporal spread to certify anomalous diffusion, $0<2d<1$.
	Let us begin with the former and then proceed with the latter.

	For a fixed step number $t$, and for $p=0$, an exponentially localized distribution is expected, Anderson localization.
	With increasing disorder level $p$, we expect a broadening of the distribution.
	Eventually, for $p=1$, a Gaussian shape should be obtained, typical for the diffusive behavior. 
	In order to find the parameters that fit the measured distribution best, it is convenient to work with a modified expression of Eq. \eqref{eq:spatial},
	\begin{equation}
		\label{eq:probability_distribution_log}
		\ln(P) = \left(-\left|\frac{a}{\sigma}\right|^b\right) |x|^{b} -  \ln\left(\sum_x e^{-|ax/\sigma|^b}\right),
	\end{equation}
	that can be fitted to our experimental data to characterize the probability distribution.

\begin{figure}[t]
 	\includegraphics[scale=0.60]{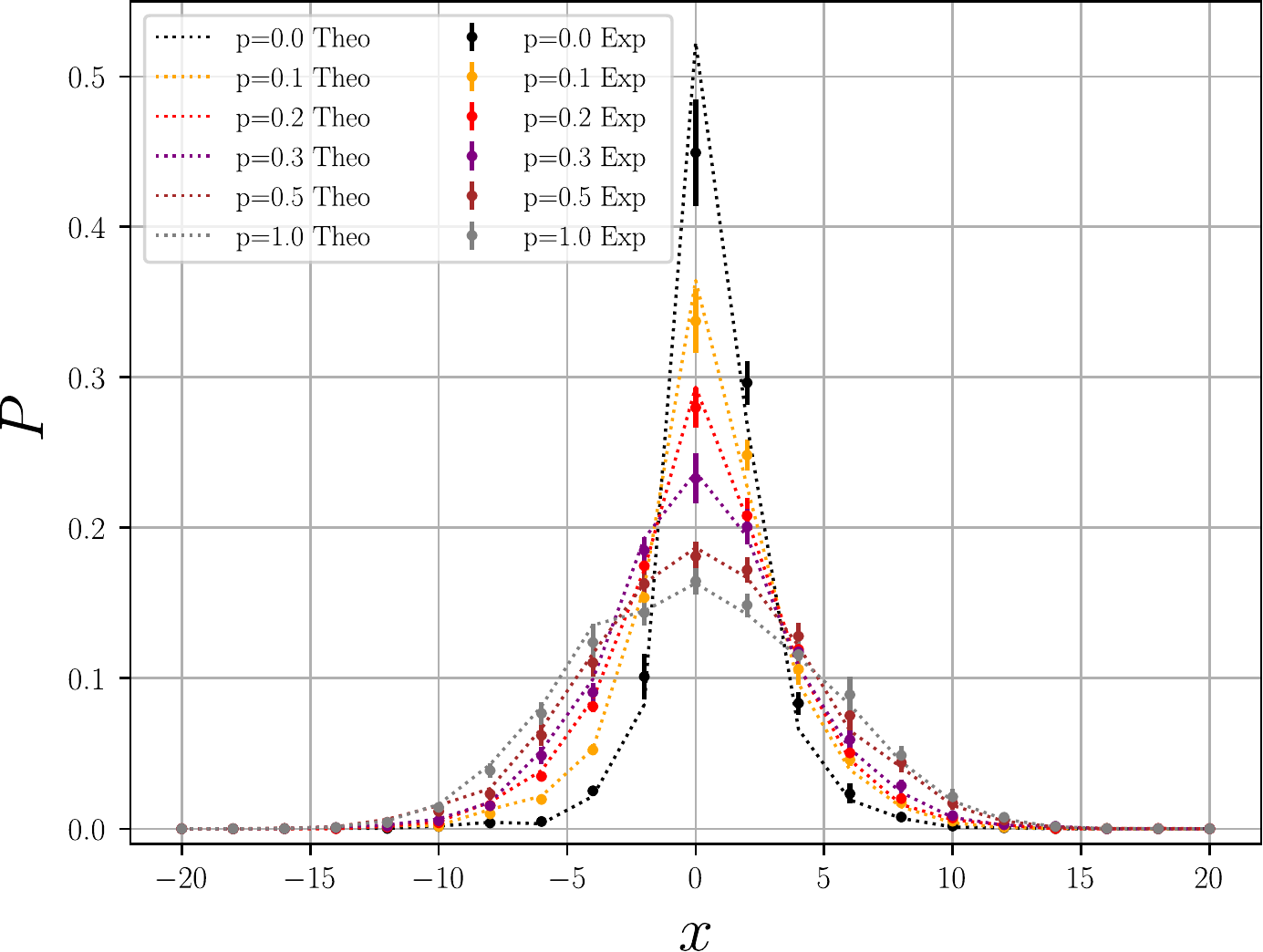}
 	\\
 	\includegraphics[scale = 0.55]{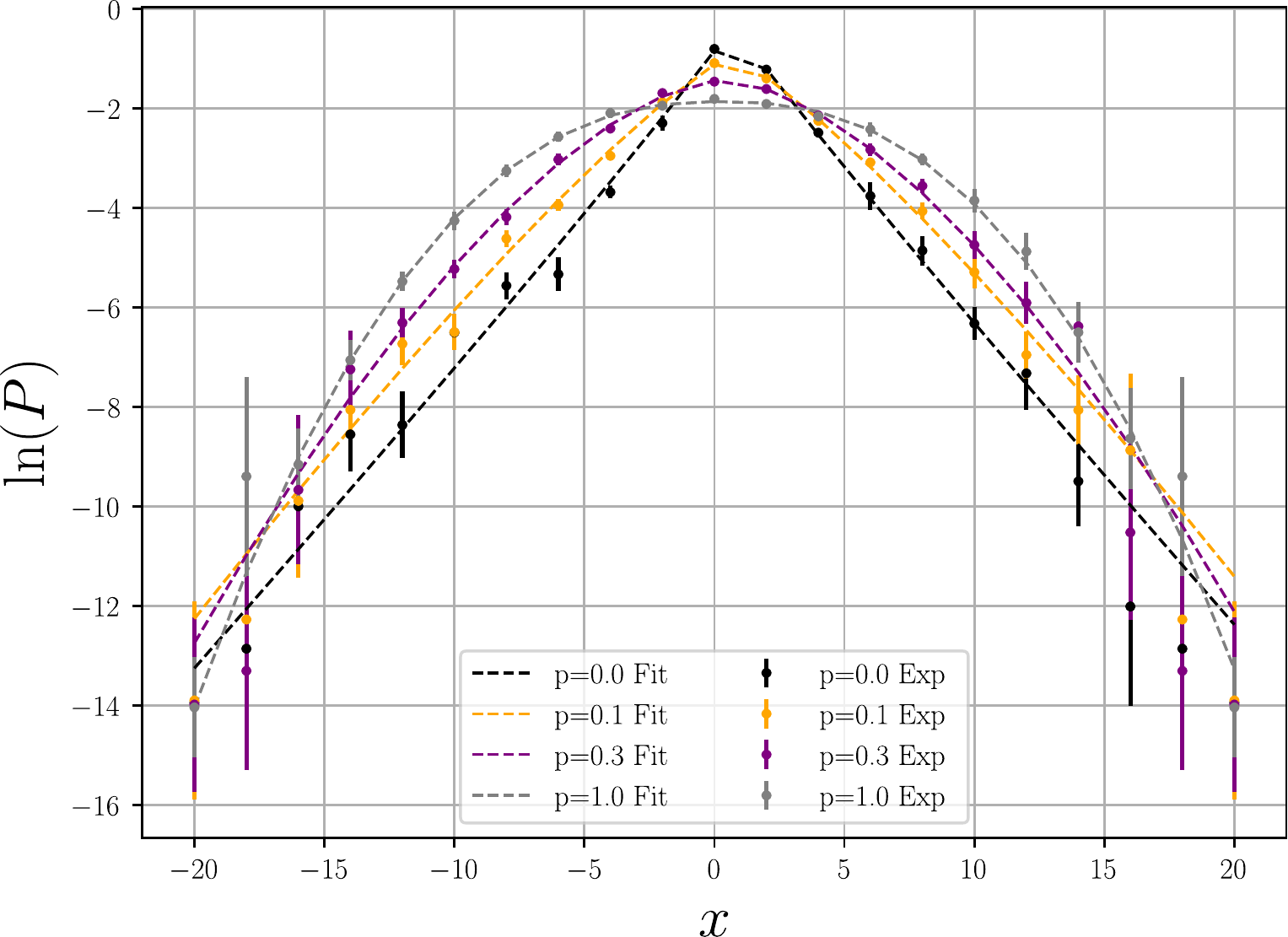}
 	\caption{
		Top.
		Probability distribution $P$ for various values of the disorder $p$, Eq. \eqref{eq:ps}.
		Experimental data (dots) agree within the uncertainties with the theoretical results (dotted lines).
		Error bars take into account Poissonian statistics and experimental imperfections of the setup.
		Bottom.
		Logarithm of the experimental probability distribution (dots) together with fit (dashed lines) according to Eq. \eqref{eq:probability_distribution_log}.
		For the sake of clarity, only selected $p$ values are depicted.
	}\label{fig:prob_step_20}
 \end{figure}
 
	Experimental data corresponding to $t=20$ for different amounts of disorder are reported in Fig. \ref{fig:prob_step_20}.
	In the top plot, dots correspond to experimentally obtained probability distributions.
	Dotted lines represent values of the theoretical probability distribution, obtained from a numerical simulation.
	Similarities between experimental and theoretical probability distributions for each step and $p$ values are above $99\%$, indicating a very good agreement even without considering many unavoidable experimental imperfections in our simulation.
	The bottom panel of Fig. \ref{fig:prob_step_20} is even more conclusive when it comes to determining the characteristic exponent $b$.
	The plot shows experimental data (dots) together with the fitted curves (dashed lines) according to  Eq. \eqref{eq:probability_distribution_log}, covering the range from a linear ($b\approx 1$) to a parabolic ($b\approx 2$) decay in this logarithmic depiction.
	It is clear from the graph that the presence of higher disorder $p$ diminishes the probability to find the walker in the starting position $x=0$ for $t>0$.
	Consequently, the probability to find it in more distant positions increases, resulting in a broadened distribution.
	The subdiffusivity of the evolution of the walker is confirmed through $1\lesssim b \lesssim 2$ (the specific numerical values in Table \ref{tab:fits}).
	It is worth noting that other imperfections lead to a broader range than one actually expects from an ideal model, cf. value $b>2$ in Table \ref{tab:fits}.

\begin{table}[t]
	\caption{
		Values of the characteristic exponents for the spatial (values for $b$) and temporal (values for $2d$) behavior of $p$-diluted subdiffusive QWs.
		By collecting data from realizing $400$ coin maps, we fit the theoretical predictions in Eqs. \eqref{eq:probability_distribution_log} and \eqref{eq:variance_subdiffusion_linearized} to the measured data in Figs. \ref{fig:prob_step_20} (bottom) and \ref{fig:var_vs_step_1photon}.
	}\label{tab:fits}
	\begin{tabular}{ccccc}
		\hline\hline
		$p$ & \hspace*{1.75cm} & $b$ & \hspace*{1.75cm} & $2d$
		\\
		\hline
		$0.0$ && $0.953 \pm 0.044$ && $0.346 \pm 0.040$
		\\
		$0.1$ && $1.199 \pm 0.048$ && $0.551 \pm 0.030$
		\\
		$0.2$ && $1.489 \pm 0.046$ && $0.723 \pm 0.032$
		\\
		$0.3$ && $1.639 \pm 0.066$ && $0.812 \pm 0.028$
		\\
		$0.5$ && $2.071 \pm 0.077$ && $0.947 \pm 0.027$
		\\
		$1.0$ && $2.422 \pm 0.083$ && $1.070 \pm 0.032$
		\\
		\hline\hline
	\end{tabular}
\end{table}

	The second feature we focus on consists of the dependency of the variance as a function of the step number $t$, again for different values of disorder.
	To assess the subdiffusive spread with our data, it is similarly convenient to recast  Eq. \eqref{eq:temporal} into a logarithmic form,
	\begin{equation}
		\label{eq:variance_subdiffusion_linearized}
		\ln(\sigma^2) = 2d\ln(t) +\ln(c^2).
	\end{equation}

\begin{figure}[b]
 	\includegraphics[scale=0.60]{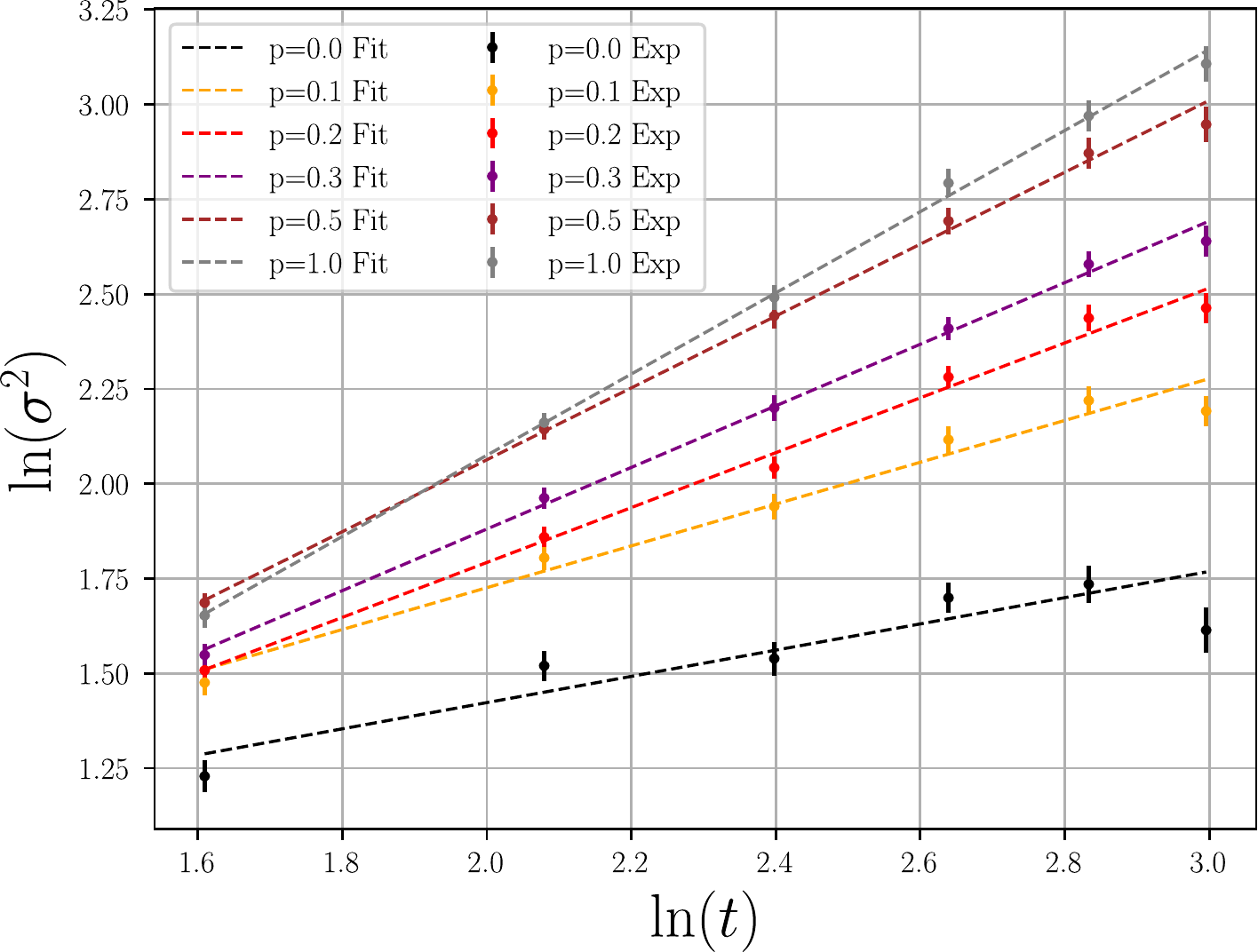}
 	\caption{
		Logarithm of the variance as a function of the logarithm of the step number $t$ [Eq. \eqref{eq:ts}] for disorder values $p$ in Eq. \eqref{eq:ps}.
		Dots correspond to experimental data; dashed lines show fits according to Eq. \eqref{eq:variance_subdiffusion_linearized}.
		The linear behavior with slopes between zero and one in this doubly logarithmic graph for each value of $p$ demonstrates an excellent agreement with the predicted subdiffusive nature of the evolution.
	}\label{fig:var_vs_step_1photon}
\end{figure}

	Results are reported in Fig. \ref{fig:var_vs_step_1photon} on a logarithmic scale for both axes.
	Dashed lines correspond to the curve in Eq. \eqref{eq:variance_subdiffusion_linearized} which is fitted to the experimental data (dots) for different values of the disorder $p$.
	The nearly perfect linear behavior with slopes $0\lesssim 2d\lesssim 1$ (see Table \ref{tab:fits} for numerical values) confirms the actual subdiffusive spread of the QW evolution.
	A discrepancy between the data and the fit can be observed for $p=0.0$ because of the extreme sensitivity of Anderson localization with respect to unavoidable experimental imperfections.
	Here, one would also expect a constant variance, which, however, is only approached in the limit $t\to\infty$, even in theory.
	As in the previous spatial analysis, error bars on the experimental data have been computed considering a Poissonian statistic of counting as well as experimental imperfections of the setup.

	Beyond earlier studies, we analyzed both the spatial and temporal impact of the amount of disorder $p$.
	According to our results, we can confirm that our approach enables us to simulate almost any subdiffusive behavior.
	Thus, the sublinear spread of the walker over time and the characteristic shapes of the measured spatial distributions indicate that the interplay between a static disorder and completely random disorder, freely controlled and interpolated via $p$, is a viable way to reproduce complex subdiffusion phenomena through discrete QWs.

\paragraph*{Conclusion.---}

	In pursuing the ultimate goal of implementing a universal quantum simulator, we demonstrated the ability to experimentally simulate subdiffusive transport phenomena, having a wide range of applications, via disordered QWs.
	By analyzing our data regarding their spatial and temporal features, we have been able to map the landscape of characteristic properties of subdiffusion.
	Firstly, we controlled our system in a way that leads to position distributions of the walker ranging from Anderson localization to a normal Gaussian distribution.
	Secondly, anomalous diffusion in the sublinear regime was explored to characterize the spread of the walker over time.
	This complements earlier findings that have been restricted to superdiffusion by starting from an already completely ordered evolution.

	Because of our unique control over the coin at each position (time bin) and for each step of the QW, the demanding goal of realizing subdiffusion was successfully accomplished with our setup.
	By perturbing our initial implementation of static disorder, we realized a $p$-diluted QW by adding dynamic noise in a controlled manner to steer our system towards the subdiffusive regime.
	Specifically, this method introduces additional fluctuations, with probabilities $p$ and $1-p$ for the dynamic and the static disorder contributions, respectively.
	The agreement between the measured data and the theoretical predictions for both the quantities under study, namely shape of the position distribution and the change of the variance in time, clearly demonstrates that the coherent walker evolves subdiffusively.

	Exceeding our proof-of-concept realization reported here, our results provide a promising starting point for future studies as well.
	For instance, our setup actually enables us to measure coin-space-resolved distributions (see SM) that can be relevant for assessing quantum properties between the coin and time-bin degrees of freedom, such as entanglement.
	Furthermore, the experiment could be extended to two single-photon walkers \cite{geraldi2019superdiffusion} by means of the very same experimental setup \cite{Nitsche2020localvsglobal}.
	This could further foster other simulations of sophisticated correlated diffusion phenomena.
	For instance, it is known that Anderson localization holds true in the case of entangled photons \cite{crespi2013anderson}.
	However, the general impact of correlated $(p_1,p_2)$-diluted dynamical noise for the walkers $1$ and $2$ is entirely unknown but could potentially be studied in our system.

\paragraph*{Acknowledgments.---}

	The Integrated Quantum Optics group acknowledges financial support through the Gott\-fried Wilhelm Leibniz-Preis (Grant No. SI1115/3-1) and the European Commission through the ERC project QuPoPCoRN (Grant No. 725366).
	A. G., A. L., and P. M. acknowledge support from the European Commission grants FP7-ICT-2011-9-600838 (QWAD - Quantum Waveguides Application and Development).

\bibliography{apssamp}


\onecolumngrid
\clearpage
\appendix*
\section{SUPPLEMENTAL MATERIAL}
\twocolumngrid

\subsection{Detailed description experimental setup}

\begin{figure}[b]
	\includegraphics[width=\columnwidth]{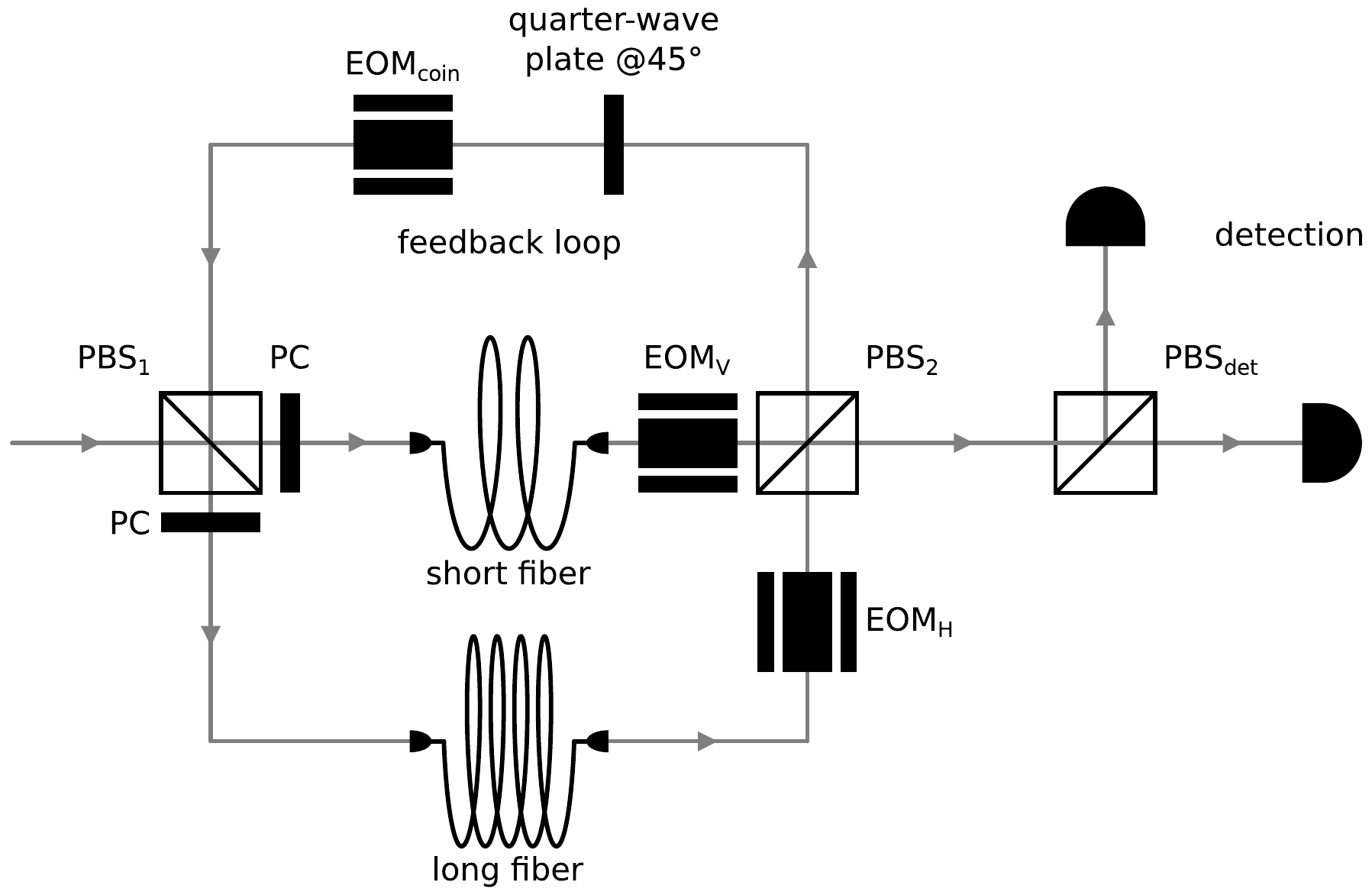}
	\caption{
		Schematics of the experimental layout, using the following elements:
		polarization controller (PC),
		polarizing beam splitter (PBS),
		and electro-optical modulator (EOM).
		The PC allows to precisely compensate the polarization rotation caused by the propagation through the fibers.
		Note that the detection is, in fact, polarization resolving.
		This is achieved by splitting the output of the loop with $\textrm{PBS}_\textrm{det}$ followed by one detector for each polarization.
	}\label{fig:experimental_setup}
\end{figure}

	Here we present details of our time-multiplexing QW setup based on a fiber loop as shown in Fig. \ref{fig:experimental_setup}.
	See also Refs. \cite{schreiber2010photons, schreiber2011decoherence, Nitsche2018recurrence}.
	This scheme is beneficial in terms of resource efficiency, high stability, and high homogeneity, which we exploit to realize coherent evolution over sufficiently large number of steps.

	A coherent laser pulse, attenuated to a single-photon per pulse on average, serves as the quantum walker.
	The walker pulse is derived from a laser with central wavelength of $1550~\textrm{nm}$, pulse width of $1~\textrm{ps}$, and repetition rate of $\sim 4~\textrm{kHz}$.
	The walk starts when the pulse impinges from the top port of $\textrm{PBS}_1$ for the first time.
	The walk is then initialized at position $x=0$ with horizontally polarized light, $|\psi(0)\rangle=|0\rangle\otimes|0\rangle$.
	The unbalanced Mach-Zehnder interferometer implements the step operation $\hat{S}$, which includes polarization dependent splitting at $\textrm{PBS}_1$, propagation of horizontal and vertical polarization through long ($\sim 473~\textrm{m}$) and short ($\sim 453~\textrm{m}$) fibers, respectively, and finally coherent recombination of the two paths at $\textrm{PBS}_2$ to introduce a well-defined delay between the two polarizations.
	The interferometer is closed with a free-space feedback loop, which redirects the light back to $\textrm{PBS}_1$ for the next step.
	In totality, this leads to the encoding of walker's position degree of freedom in the pulse arrival time.
	The capability of dynamical polarization rotation by the two fast-switching EOMs, $\textrm{EOM}_\textrm{H}$ and $\textrm{EOM}_\textrm{V}$, enables us routing the pulses either back to the feedback loop or to the detection unit. This high-quality active polarization control facilitates deterministic in- and out-coupling, rendering it possible to implement sufficiently large number of steps by enhancing the roundtrip efficiency.
	The current setup is designed to have a step separation of $\sim 2.3~\mu\textrm{s}$ and position separation of $\sim 105~\textrm{ns}$ and has been utilized to demonstrate walks up to 36 steps by allowing time-bin interlacing for successive steps \cite{Nitsche2018recurrence}.
	However, we here restrict ourselves here to 20 steps, which is sufficient to unambiguously discern subdiffusive dynamics, while minimizing the error from interlacing.
	The detection unit allows for polarization-resolved photon counting at individual time bins, using  $\textrm{PBS}_\textrm{det}$ and high-efficiency ($> 90\%$) superconducting nanowire single-photon detectors with a dead time of $\sim 100~\textrm{ns}$, from which we deduce the evolution of walker's probability distributions.

	Our investigation of subdiffusive behavior mainly relies on the implementation of position and step dependent coin operation, $\hat{C}(x,t)$.
	This dynamical coin operation is achieved by extending the capability of the previous setup via the introduction of another fast-swithing EOM ($\textrm{EOM}_\textrm{coin}$) followed by a quarter-wave plate (QWP) in the feedback path.
	The action of a QWP aligned at an angle $45^\circ$ with respect to the polarization basis $\{|H\rangle,|V\rangle\}$ reads as 
	\begin{equation}
		\hat{C}_{\textrm{QWP}}=\frac{1}{\sqrt{2}}
		\begin{pmatrix}
			1 & -i \\
			-i & 1
		\end{pmatrix}.
	\end{equation}
	The EOM operation can be written as
	\begin{equation}
		\hat{C}_{\textrm{EOM}}=
		\begin{pmatrix}
			\cos{\phi} & -i\,\sin{\phi} \\
			-i\,\sin{\phi} & \cos{\phi}
		\end{pmatrix},
	\end{equation}
	where the phase $\phi$ can be tuned by varying the voltage applied to the EOM.
	Their combination leads to the transformation
	\begin{equation}
		\hat{C}_{\textrm{EOM}}\, \hat{C}_{\textrm{QWP}}=
		\begin{pmatrix}
			\cos{\theta} & -i\,\sin{\theta} \\
			-i\,\sin{\theta} & \cos{\theta}
		\end{pmatrix},
	\end{equation}
	using
	\begin{equation}
		\theta=\phi+\frac{\pi}{4}
	\end{equation}
	and the identities $(\cos{\phi} - \sin{\phi})/\sqrt 2 = \cos{\theta}$ and $(\cos{\phi} + \sin{\phi})/\sqrt 2 = \sin{\theta}$. 

	It is worth emphasizing that the present scheme utilizes free-space EOMs, which introduce very low losses ($< 1\%$).
	The combination of active in- and out-coupling and free-space EOMs leads to a significantly improved roundtrip efficiency ($>80\%$) in comparison to the previous disordered time-multiplexing QW setup that used integrated EOM \cite{schreiber2011decoherence}.
	However, relatively high-voltage requirements for free-space EOMs comes with the hardware limitations that allow only three different voltage settings, $v\in\{-v_1,0,+ v_1\}$, during a single experimental run.
	In particular, $v=0$ corresponds to $\phi = 0$, leading to a coin operation that equally mixes $|H\rangle$ and $|V\rangle$.
	We chose $v=\pm v_1$ such that $\phi=\mp\pi/4$.
	This yields an identity coin that leaves the polarization states unchanged and a reflection coin that switches the polarizations.
	Notably, we find that these three accessible coin operations are sufficient for the exploration of the complete subdiffusive QW regime, thanks to the $p$-diluted disorder scheme.
	We design appropriate voltage-switching patterns for the EOMs that put into effect various disorder strengths $p$ ranging from $p=0$ (Anderson localization) to $p=1$ (normal diffusion). 

\subsection{Supplemental details on the theory}

	For a self-consistent reading of this work and for a coherent treatment, we recapitulate and reformulate the theory on diffusion in randomized media as reported in Ref. \cite{giona1992fractional}.
	There, the approach was based on the Laplace transform in the temporal domain.
	For our purposes, it is, however, more convenient to discuss that method in terms of the Fourier transform in the spatial domain.
	Eventually, we relate this approach to $p$-diluted models.

	A general model of diffusion in a one-dimensional system can be described by the equation
	\begin{equation}
		\label{appeq:GenHeatEq}
		0=\partial_tP(x,t)+\mathcal L(-\partial_x^2,t)P(x,t),
	\end{equation}
	where $\mathcal L$ is a potentially time-dependent differential operator.
	In the continuous limit, this equation also models the asymptotic behavior of a discrete system, such as our 
	QW.
	Furthermore, the differential operator depends on $-\partial_x^2$ for a positive (i.e., dispersive) behavior because of $-\partial_x^2 e^{ixk}=k^2e^{ixk}$ and $k^2\geq0$.

	Using the characteristic function, i.e., the Fourier transform $\Phi(k,t)=\int_{-\infty}^{+\infty} dx\, e^{-ikx} P(x,t)$, we can rewrite Eq. \eqref{appeq:GenHeatEq} as
	\begin{equation}
		\label{appeq:GenHeatEqFT}
		0=\partial_t\Phi(k,t)+\mathcal L(k^2,t)\Phi(k,t).
	\end{equation}
	Then, the solution in form of the Green's function can be formally expressed as
	\begin{equation}
		\label{appeq:GreensFct}
		\tilde G(k,t)=\exp\left(-\int_0^tdt'\,\mathcal L(k^2,t')\right),
	\end{equation}
	This solves Eq. \eqref{appeq:GenHeatEqFT} as $\Phi(k,t)=\tilde G(k,t)\Phi(k,0)$, where $\Phi(k,0)$ represents the inital distribution.
	In our case, this is modeled by a singular input at the center position, thus $\Phi(k,0)=1$ in the Fourier domain.

	It was also shown in Ref. \cite{giona1992fractional} that, for large times ($t\gg1$), solutions follow the functional form
	\begin{equation}
		\label{appeq:ExpDist}
		P(x,t)=\frac{ab}{2\sigma(t)\Gamma(1/b)}\exp\left(
			-\left|\frac{ax}{\sigma(t)}\right|^b
		\right),
	\end{equation}
	with $\Gamma$ being the Gamma function and $\sigma(t)$ denoting a time-dependent standard deviation.
	In addition, we define $a=\sqrt{\Gamma(3/b)/\Gamma(1/b)}$ and $b$ relates to the type of exponential decay;
	e.g., $b=1$ and $b=2$ define a linear and quadratic behavior, respectively.

	The moments of this distribution can be evaluated as well;
	odd moments vanish and even moments read
	\begin{equation}
		E(x^{2n})=
		\frac{\Gamma\left(\frac{2n+1}{b}\right)}{\Gamma\left(\frac{1}{b}\right)}
		\left[\frac{\sigma(t)}{a}\right]^{2n}.
	\end{equation}
	These moments enable us to expand the characteristic function in a Taylor series as
	\begin{equation}
		\label{appeq:CharFctTaylor}
	\begin{aligned}
		&\Phi(k,t)
		=\sum_{n=0}^\infty E(x^n) \frac{[ik]^n}{n!}
		\\
		=&1
		+\sigma(t)^2\frac{-k^2}{2}
		+\frac{\Gamma\left(\frac{5}{b}\right)\Gamma\left(\frac{1}{b}\right)}{\Gamma\left(\frac{3}{b}\right)^2}
		\sigma(t)^{4}\frac{k^4}{24}
		+\cdots.
	\end{aligned}
	\end{equation}

\begin{figure}[b]
	\includegraphics[width=.8\columnwidth]{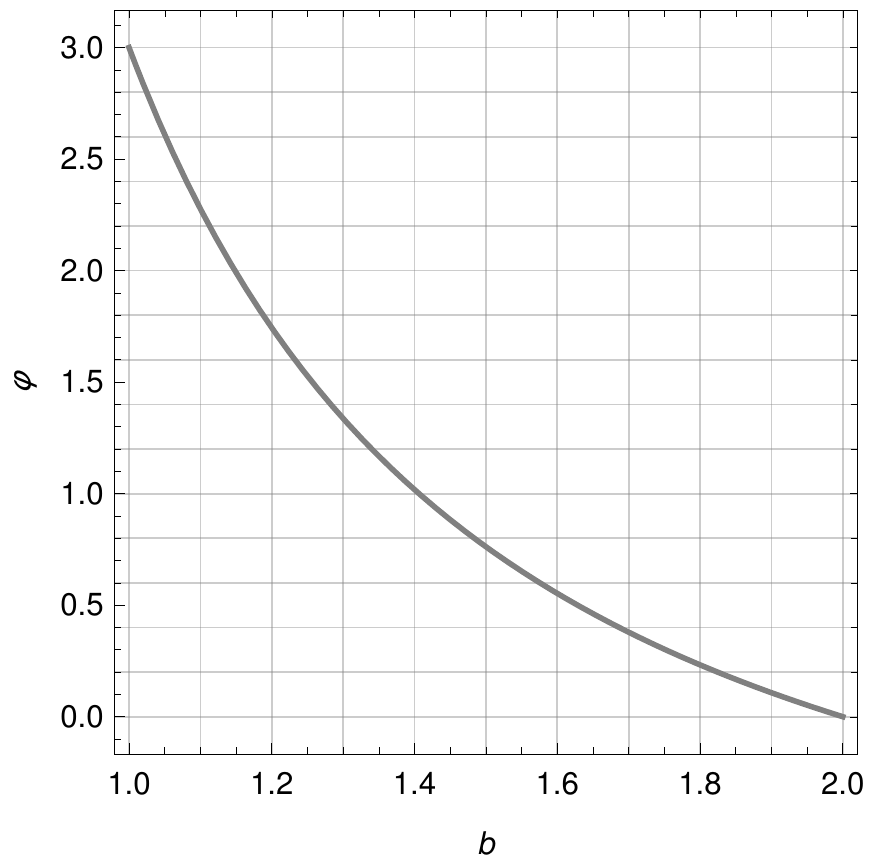}
	\caption{
		Function $\varphi=f(b)=\Gamma\left(\frac{5}{b}\right)\Gamma\left(\frac{1}{b}\right) / \Gamma\left(\frac{3}{b}\right)^2-3$ is shown in the relevant interval $1\leq b\leq 2$.
		In this region, $f$ is strictly monotonously decreasing, allowing for defining its inverse for determining $b$ from $\varphi$.
	}\label{appfig:fcalibration}
\end{figure}

	Similarly, we can expand the generator of the evolution, $\mathcal L(k^2,t)=\sum_{n=0}^\infty\lambda_{2n}(t)\frac{k^{2n}}{(2n)!}$.
	This further allows us to expand the Green's function from Eq. \eqref{appeq:GreensFct},
	\begin{equation}
	    \label{appeq:GreensFctTaylor}
	\begin{aligned}
		&\tilde G(k,t)
		\\
		=&\tilde G(0,t)
		+\left[{-}\int_0^tdt'\,\lambda_2(t')\right]\tilde G(0,t)\frac{k^2}{2}
		\\
		&+\left[
			{-}\int_0^tdt'\,\lambda_4(t')
			+3\left[\int_0^tdt'\,\lambda_2(t')\right]^2
		\right]\tilde G(0,t)\frac{k^4}{24}
		\\
		&+\cdots,
	\end{aligned}
	\end{equation}
	where $\tilde G(0,t)=\exp\left(\int_0^tdt'\,\lambda_0(t')\right)$.
	Because of our initial conditions, resulting in $\Phi(k,t)=\tilde G(k,t)$, we can now equate the coefficients for $\tilde G$ in Eq. \eqref{appeq:GreensFctTaylor} and $\Phi$ in Eq. \eqref{appeq:CharFctTaylor}.
	Since this identification has to be satisfied for all times $t>0$, we find
	\begin{equation}
	    \label{appeq:MeanCoeff}
	\begin{aligned}
		\lambda_0(t) = 0,\quad
	    \int_0^tdt'\,\lambda_2(t') = \sigma(t)^2,\quad
		\text{and}
		\\
		-\int_0^tdt'\,\lambda_4(t') = \underbrace{\left[
			\frac{\Gamma\left(\frac{5}{b}\right)\Gamma\left(\frac{1}{b}\right)}{\Gamma\left(\frac{3}{b}\right)^2}-3
		\right]}_{\stackrel{\text{def.}}{=}\varphi=f(b)} \sigma(t)^4.
	\end{aligned}
	\end{equation}
	Importantly, $b=f^{-1}(\varphi)$ determines the exponent in Eq. \eqref{appeq:ExpDist}.
	See Fig. \ref{appfig:fcalibration} for the graph of $f$.

    A first consequence of the aforementioned relations is that the spread $\sigma(t)$ is given by the time dependency of the first nonzero Taylor coefficient $\lambda_{2}(t)$, resulting the the corresponding power law, such as $\sigma(t)=ct^d$ with constants $c,d>0$ \cite{giona1992fractional}.
    Thus, the introduction dynamic disorder, changing the time-dependence  generator $\mathcal L$, generally results in an increment of the power.
    As a second observation, we have a look at the relation that includes $\lambda_{4}(t)$.
    This coefficient is typically negative which allows one to substitute it by $-\lambda_4(t)=\rho(t)\lambda_2(t)^2$.
    With this, we can rewrite the above relation as
    \begin{equation}
    \begin{aligned}
        \varphi=&\frac{
            \int_0^{t}dt'\,\rho(t')\lambda_2(t')^2
            -\left(\int_0^{t}dt'\,\rho(t')\lambda_2(t')\right)^2
        }{\left(\int_0^{t}dt'\,\lambda_2(t')\right)^2}
        \\
        &+\left(\frac{\int_0^{t}dt'\,\rho(t')\lambda_2(t')}{\int_0^{t}dt'\,\lambda_2(t')}\right)^2.
    \end{aligned}
    \end{equation}
    Herein, the numerator of the first term plays a role a variance, quantifying the fluctuation in $\lambda_2(t)$, that influences $b=f^{-1}(\varphi)$.

    With these considerations, we can conclude that our $p$-diluted model dynamically changes the generator $\mathcal L$ by altering the coin operations.
    As discussed above, this broadens the spread in time (increasing $d$ via $\lambda_2$).
    Secondly, it changes the spatial exponential decay.
    That is, if only a few coins are changed per time step (low $p$), those are unlikely the same coins, leading to a high fluctuation in $\lambda_4$, thus high $\varphi$, thus low $b$ (Fig. \ref{appfig:fcalibration}).
    The other way around, a high $p$ results in low $\varphi$ and a high $b$.
    
\subsection{Additional results from data analysis and comparison with numerical model}

\begin{figure}
 	\includegraphics[scale=0.73]{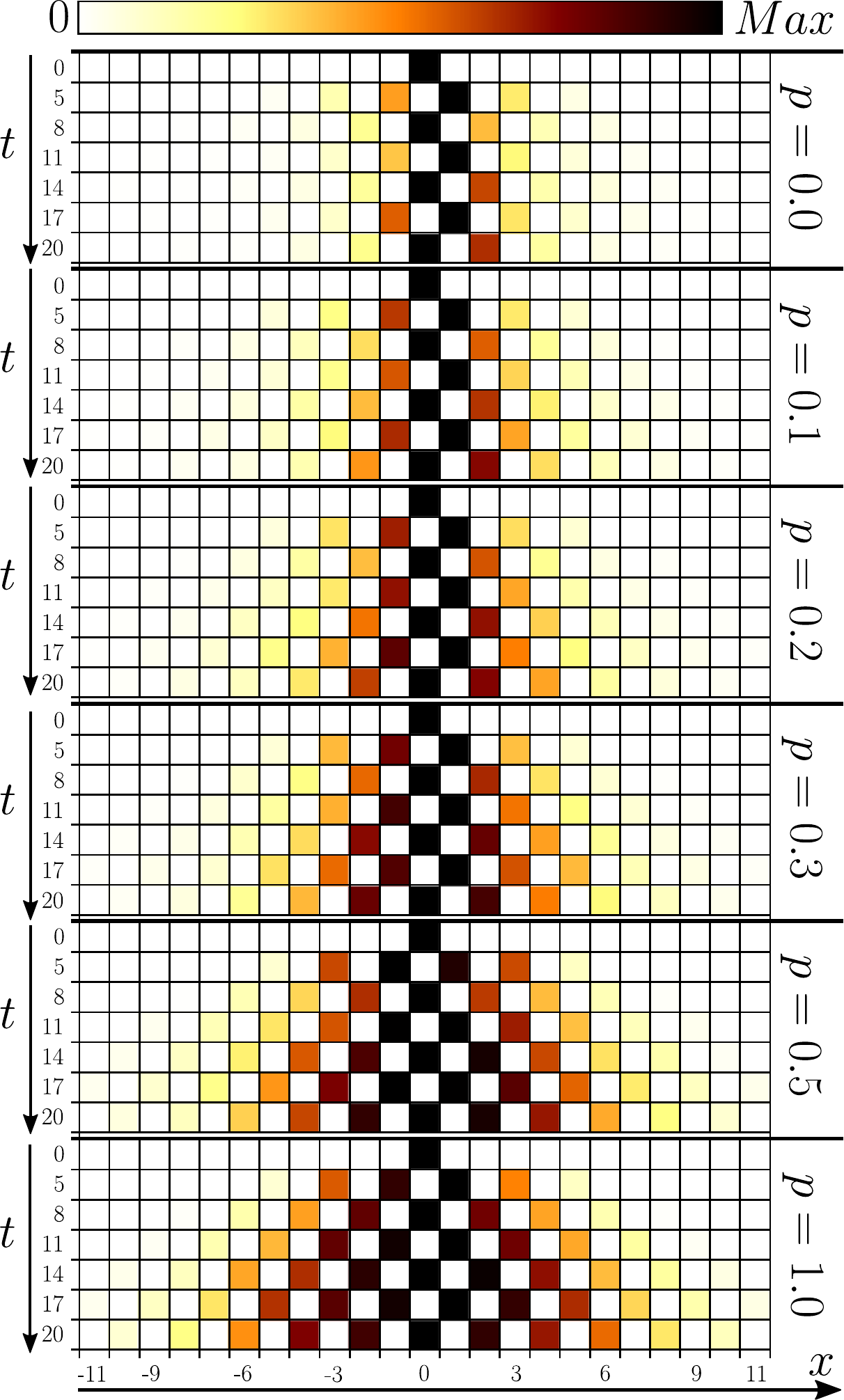}
 	\caption{
		Experimentally measured normalized intensity distributions $P(x,t)$.
		Each panel corresponds to one parameter $p$, increasing from top to bottom.
		Data are reported for $x\in[-11,11]$ (horizontal axis) and for several selected times $t\in[5,20]$ (vertical axis).
	}\label{fig:evolution_total}
\end{figure}

\begin{figure}
 	\includegraphics[scale=0.61]{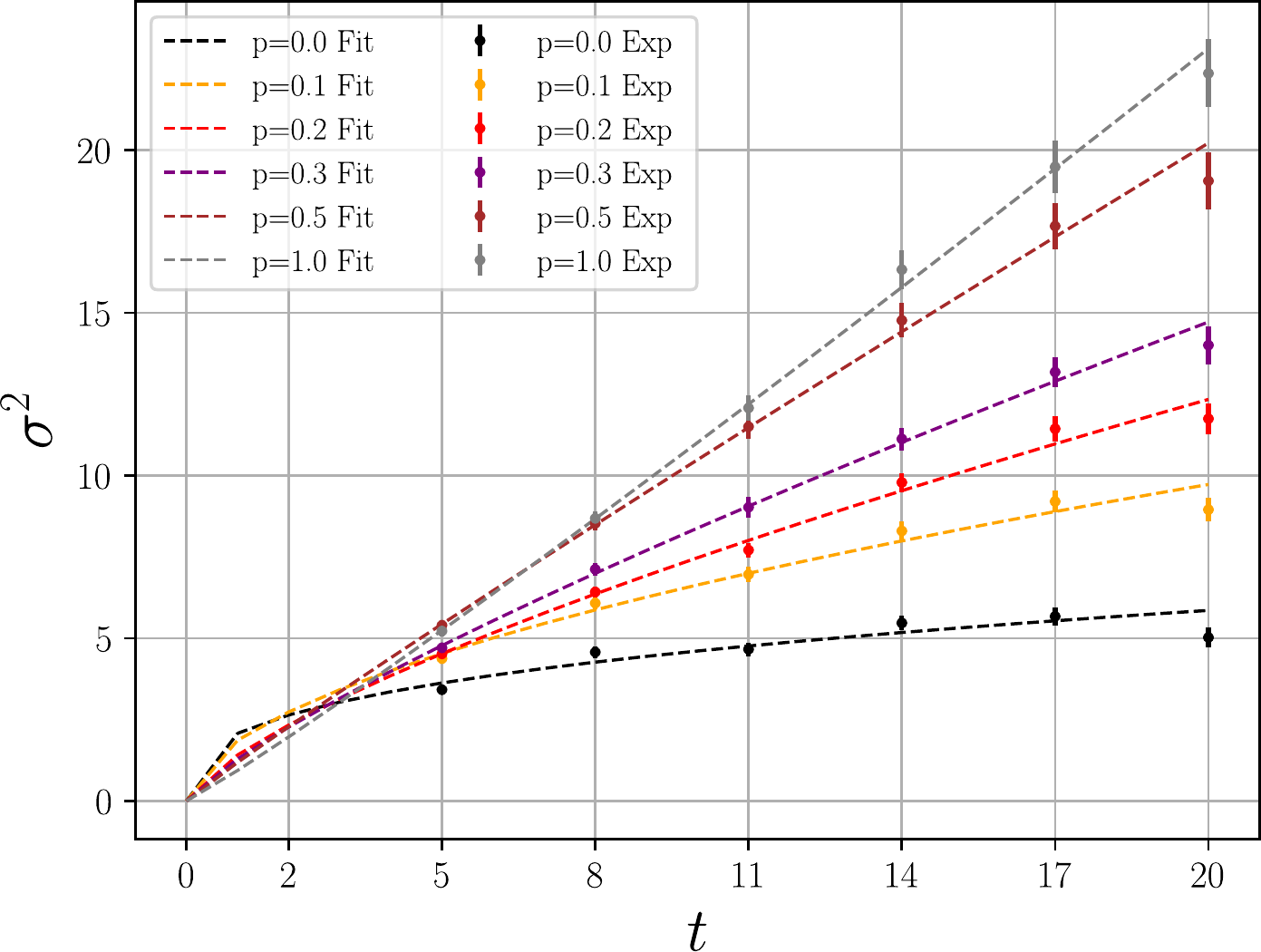}
 	\caption{
		Experimental data (dots) and the fitted relation $\sigma^2=c^2t^{2d}$ (dashed lines) on a linear scale for both axes.
	}\label{fig:linear_variance}
\end{figure}

\begin{table*}
	\caption{
		Characteristic parameters for the implemented disorder levels $p$.
		The values have been estimated by means of a least square fit to theoretical predictions.
		Quantities with the subscript ``num'' have been extracted from our numerical model, considering $10\,000$ coin maps.
		The subscript ``exp'' indicates quantities obtained from our data, realizing $400$ coin maps.
		The parameters $b$ and $\delta$ relate to the theoretical prediction $\ln(P)=-\delta|x|^b+\mathrm{const.}$ (where $\delta=|a/\sigma|^b$ when compared to main text) for the 20th step.
		Values for $2d$ and $c^2$ are a result of the fit to the theoretical prediction $\ln(\sigma^2)=2d\ln(t)+\ln(c^2)$.
	}\label{tab:expvnum}
	\begin{tabular}{
		p{.07\textwidth}
		p{.07\textwidth}p{.16\textwidth}
		p{.07\textwidth}p{.16\textwidth}
		p{.07\textwidth}p{.16\textwidth}
		p{.07\textwidth}p{.1\textwidth}
	}
		\hline\hline
		$p$ & $b_\mathrm{num}$ & $b_\mathrm{exp}$ & $\delta_\mathrm{num}$ & $\delta_\mathrm{exp}$ & $2d_\mathrm{num}$ & $2d_\mathrm{exp}$ & $c^2_\mathrm{num}$ & $c^2_\mathrm{exp}$
		\\
		\hline
		$0.0$ & $0.800$ & $0.953 \pm 0.044$ & $1.027$ & $0.719 \pm 0.084$ & $0.097$ & $0.346 \pm 0.040$ & $3.56$ & $2.08\phantom{0} \pm 0.19\phantom{0}$
		\\
		$0.1$ & $1.126$ & $1.199 \pm 0.048$ & $0.367$ & $0.300 \pm 0.034$ & $0.504$ & $0.551 \pm 0.030$ & $1.88$ & $1.87\phantom{0} \pm 0.14\phantom{0}$
		\\
		$0.2$ & $1.378$ & $1.489 \pm 0.046$ & $0.171$ & $0.130 \pm 0.015$ & $0.686$ & $0.723 \pm 0.032$ & $1.44$ & $1.41\phantom{0} \pm 0.11\phantom{0}$
		\\
		$0.3$ & $1.568$ & $1.639 \pm 0.066$ & $0.095$ & $0.081 \pm 0.013$ & $0.776$ & $0.812 \pm 0.028$ & $1.376$ & $1.293 \pm 0.087$
		\\
		$0.5$ & $1.863$ & $2.071 \pm 0.077$ & $0.038$ & $0.022 \pm 0.004$ & $0.894$ & $0.947 \pm 0.027$ & $1.232$ & $1.183 \pm 0.073$
		\\
		$1.0$ & $2.138$ & $2.422 \pm 0.083$ & $0.016$ & $0.008 \pm 0.002$ & $1.043$ & $1.070 \pm 0.032$ & $0.967$ & $0.937 \pm 0.070$
		\\
		\hline\hline
	\end{tabular}
\end{table*}

	Experimental probability distributions as a function of $x$ and $t$ are reported in Fig. \ref{fig:evolution_total} for all selected $p$ values, showing positions $-11<x<11$ for all selected time steps $t$.
	For an enhanced data visualization, each row is normalized to the maximum of the corresponding probability distribution.
	It is clear that the spread of the distribution increases with $p$, starting from a condition in which the walker remains localized for all the steps of the evolution ($p=0$) up to the behavior typical of a completely disordered QW ($p=1$).
	In addition to the logarithmic plot of the time-dependent variance in the main text, a linearly scaled version is provided in Fig. \ref{fig:linear_variance}.

	In order to compare our experimental results with the expected ones, we implemented a numerical simulation that produces $10\,000$ different coin maps for a given level of disorder.
	Let us recall that a coin map is a set of coin configurations which are obtained by starting from static disorder $\hat C(x)$ and randomly changing a percentage $p$ of coins to $\hat C(x,t)$.
	The distribution that is obtained by averaging over all numerically implemented coin maps then models our experiment.
	However, these theoretical values have been computed in the completely ideal case, i.e., without considering unavoidable setup imperfections.
	Still, this simple model was already sufficient to match the results of the experiment sufficiently well.

	In addition to the space-time dependent depiction of our data in Fig. \ref{fig:evolution_total}, the comparison to theory of both our data and numerical model are given in Table \ref{tab:expvnum} for various cases of $p$-diluted disorder.
	These values follow the expected trend:
	the higher the disorder, the higher the exponential decay in space and temporal dispersion, quantified by $b$ and $d$, respectively.
	Specifically, the reported values confirm that we mostly operate in the subdiffusive regime of the QW, $1\leq b\leq 2$ and $0\leq 2d\leq 1$.

	As one might expects, small discrepancies can be observed between experimental and numerical values as well as the theoretical predictions.
	For instance, deviations from numerical and experimental parameters can be caused by imperfect randomization since these parameters have been extracted by averaging the probability distributions over $10\,000$ coin maps in the ideal simulation while only $400$ have been implemented experimentally.
	Nevertheless, estimates for parameters from numerics and data mostly agree with each other within the confidence interval, and deviations can be generally explained by considering experimental imperfections, such as a nonideal operation of the EOMs as well as the QWP, less than $100\%$ visibility of interference, and slight setup misalignment, all of which contribute to increasing the spread of the walker.

	The highest discrepancies to the theory can be observed for the extremal cases of disorder, $p=0$ and $p=1$, affecting both numerics and experiment.
	Firstly, the discrepancy for $p\to 0$ can be understood by considering that Anderson localization typically arises from a strict periodicity in the disorder pattern.
	For this reason, it is much more sensitive to small imperfections compared to other disorder values, resulting in an higher deviation from the theory.
	Moreover, a nonspreading regime is only feasible for $t\to\infty$.
	Secondly, the discrepancy for $p\to 1$ is amplified by some of the effects previously mentioned even further.
	For instance, $p$-diluted models describe a convolution of the initial (Anderson-like) behavior with another distribution for dynamic disorder, causing that imperfections of all initial realizations add up.
	Besides, imperfections propagate along with the spreading of the walker over time.

\end{document}